\documentstyle[twoside,graphicx]{article}

\catcode`\@=11
\long\def\@makefntext#1{
\protect\noindent \hbox to 3.2pt {\hskip-.9pt  
$^{{\eightrm\@thefnmark}}$\hfil}#1\hfill}		%CAN BE USED 

\def\@makefnmark{\hbox to 0pt{$^{\@thefnmark}$\hss}}	%ORIGINAL 
	
\def\ps@myheadings{\let\@mkboth\@gobbletwo
\def\@oddhead{\hbox{}
\rightmark\hfil\eightrm\thepage}   
\def\@oddfoot{}\def\@evenhead{\eightrm\thepage\hfil
\leftmark\hbox{}}\def\@evenfoot{}
\def\sectionmark##1{}\def\subsectionmark##1{}}

\oddsidemargin=\evensidemargin
\newcounter{sectionc}\newcounter{subsectionc}\newcounter{subsubsectionc}
\renewcommand{\section}[1] {\vspace{12pt}\addtocounter{sectionc}{1} 
\setcounter{subsectionc}{0}\setcounter{subsubsectionc}{0}\noindent 
	{\tenbf\thesectionc. #1}\par\vspace{5pt}}
\renewcommand{\subsection}[1] {\vspace{12pt}\addtocounter{subsectionc}{1} 
	\setcounter{subsubsectionc}{0}\noindent 
	{\bf\thesectionc.\thesubsectionc. {\kern1pt \bfit #1}}\par\vspace{5pt}}
\renewcommand{\subsubsection}[1] {\vspace{12pt}\addtocounter{subsubsectionc}{1}
	\noindent{\tenrm\thesectionc.\thesubsectionc.\thesubsubsectionc.
	{\kern1pt \tenit #1}}\par\vspace{5pt}}

\newcounter{appendixc}
\newcounter{subappendixc}[appendixc]
\newcounter{subsubappendixc}[subappendixc]
\renewcommand{\thesubappendixc}{\Alph{appendixc}.\arabic{subappendixc}}
\renewcommand{\thesubsubappendixc}
	{\Alph{appendixc}.\arabic{subappendixc}.\arabic{subsubappendixc}}

\renewcommand{\appendix}[1] {\vspace{12pt}
        \refstepcounter{appendixc}
        \setcounter{figure}{0}
        \setcounter{table}{0}
        \setcounter{lemma}{0}
        \setcounter{theorem}{0}
        \setcounter{corollary}{0}
        \setcounter{definition}{0}
        \setcounter{equation}{0}
        \renewcommand{\thefigure}{\Alph{appendixc}.\arabic{figure}}
        \renewcommand{\thetable}{\Alph{appendixc}.\arabic{table}}
        \renewcommand{\theappendixc}{\Alph{appendixc}}
        \renewcommand{\thelemma}{\Alph{appendixc}.\arabic{lemma}}
        \renewcommand{\thetheorem}{\Alph{appendixc}.\arabic{theorem}}
        \renewcommand{\thedefinition}{\Alph{appendixc}.\arabic{definition}}
        \renewcommand{\thecorollary}{\Alph{appendixc}.\arabic{corollary}}
        \renewcommand{\theequation}{\Alph{appendixc}.\arabic{equation}}
        \noindent{\tenbf Appendix \theappendixc #1}\par\vspace{5pt}}
\newcommand{\subappendix}[1] {\vspace{12pt}
        \refstepcounter{subappendixc}
        \noindent{\bf Appendix \thesubappendixc. {\kern1pt \bfit #1}}
	\par\vspace{5pt}}
\newcommand{\subsubappendix}[1] {\vspace{12pt}
        \refstepcounter{subsubappendixc}
        \noindent{\rm Appendix \thesubsubappendixc. {\kern1pt \tenit #1}}
	\par\vspace{5pt}}

\newcommand{\textlineskip}{\baselineskip=13pt}
\newcommand{\smalllineskip}{\baselineskip=10pt}

\def\eightcirc{
\begin{picture}(0,0)
\put(4.4,1.8){\circle{6.5}}
\end{picture}}
\def\eightcopyright{\eightcirc\kern2.7pt\hbox{\eightrm c}}

\def\abstracts#1#2#3{{
	\centering{\begin{minipage}{4.5in}\baselineskip=10pt\footnotesize
	\parindent=0pt #1\par 
	\parindent=15pt #2\par
	\parindent=15pt #3
	\end{minipage}}\par}} 

\newcommand{\bibit}{\nineit}

\renewenvironment{thebibliography}[1]
	{\frenchspacing
	 \ninerm\baselineskip=11pt
	 \begin{list}{\arabic{enumi}.}
	{\usecounter{enumi}\setlength{\parsep}{0pt}
	 \setlength{\leftmargin 12.7pt}{\rightmargin 0pt} %FOR 1--9 ITEMS
	 \setlength{\itemsep}{0pt} \settowidth
	{\labelwidth}{#1.}\sloppy}}{\end{list}}

\newcounter{itemlistc}
\newcounter{romanlistc}
\newcounter{alphlistc}
\newcounter{arabiclistc}

\newcommand{\fcaption}[1]{
        \refstepcounter{figure}
        \setbox\@tempboxa = \hbox{\footnotesize Fig.~\thefigure. #1}
        \ifdim \wd\@tempboxa > 5in
           {\begin{center}
        \parbox{5in}{\footnotesize\smalllineskip Fig.~\thefigure. #1}
            \end{center}}
        \else
             {\begin{center}
             {\footnotesize Fig.~\thefigure. #1}
              \end{center}}
        \fi}

\newcommand{\tcaption}[1]{
        \refstepcounter{table}
        \setbox\@tempboxa = \hbox{\footnotesize Table~\thetable. #1}
        \ifdim \wd\@tempboxa > 5in
           {\begin{center}
        \parbox{5in}{\footnotesize\smalllineskip Table~\thetable. #1}
            \end{center}}
        \else
             {\begin{center}
             {\footnotesize Table~\thetable. #1}
              \end{center}}
        \fi}

\def\@citex[#1]#2{\if@filesw\immediate\write\@auxout
	{\string\citation{#2}}\fi
\def\@citea{}\@cite{\@for\@citeb:=#2\do
	{\@citea\def\@citea{,}\@ifundefined
	{b@\@citeb}{{\bf ?}\@warning
	{Citation `\@citeb' on page \thepage \space undefined}}
	{\csname b@\@citeb\endcsname}}}{#1}}

\newif\if@cghi
\def\cite{\@cghitrue\@ifnextchar [{\@tempswatrue
	\@citex}{\@tempswafalse\@citex[]}}
\def\citelow{\@cghifalse\@ifnextchar [{\@tempswatrue
	\@citex}{\@tempswafalse\@citex[]}}
\def\@cite#1#2{{$\null^{#1}$\if@tempswa\typeout
	{IJCGA warning: optional citation argument 
	ignored: `#2'} \fi}}

\def\pmb#1{\setbox0=\hbox{#1}
	\kern-.025em\copy0\kern-\wd0
	\kern.05em\copy0\kern-\wd0
	\kern-.025em\raise.0433em\box0}

\def\fnt#1#2{\footnotetext{\kern-.3em
	{$^{\mbox{\scriptsize #1}}$}{#2}}}

\font\tenrm=cmr10
\font\tenit=cmti10 
\font\tenbf=cmbx10
\font\bfit=cmbxti10 at 10pt
\font\ninerm=cmr9
\font\nineit=cmti9

\font\eightrm=cmr8

\def\qed{\hbox{${\vcenter{\vbox{			%HOLLOW SQUARE
   \hrule height 0.4pt\hbox{\vrule width 0.4pt height 6pt
   \kern5pt\vrule width 0.4pt}\hrule height 0.4pt}}}$}}

\begin{document}

%\runninghead{Instructions for Typesetting Camera-Ready
%Manuscripts $\ldots$} {Instructions for Typesetting Camera-Ready
%Manuscripts $\ldots$}

\normalsize\textlineskip
\setcounter{page}{1}

%\copyrightheading{}			%{Vol. 0, No. 0 (1993) 000--000}
\vspace*{-1cm}
\begin{flushright}
ER/40685/955 \\
UB-HET-00-04\\
UR-1617 \\
November 2000
\end{flushright}

%\vspace*{0.88truein}
\vspace*{0.1truein}

%\fpage{1}
\centerline{\bf ELECTROWEAK RADIATIVE CORRECTIONS TO} 
\vspace*{0.035truein}
\centerline{\bf W AND Z BOSON PRODUCTION AT HADRON COLLIDERS\footnote{Talk 
given at the DPF 2000 meeting, Columbus, OH, August 9-12, 2000.}}
\vspace*{0.37truein}
\centerline{\footnotesize ULRICH BAUR}
\vspace*{0.015truein}
\centerline{\footnotesize\it Department of Physics, 
State University of New York at Buffalo}
\baselineskip=10pt
\centerline{\footnotesize\it Buffalo, New York 14260, USA}
\vspace*{10pt}
\centerline{\footnotesize DOREEN WACKEROTH} 
\vspace*{0.015truein}
\centerline{\footnotesize\it Department of Physics and Astronomy, 
University of Rochester}
\baselineskip=10pt
\centerline{\footnotesize\it Rochester, New York 14627, USA}
%\vspace*{0.225truein}
%\publisher{(received date)}{(revised date)}

\vspace*{0.21truein}
\abstracts{For the envisioned high precision measurement of the W boson mass
at the Tevatron and LHC it is crucial that the theoretical predictions for
the W and Z production processes are under control.  
We briefly summarize the status of the electroweak radiative corrections to
$p\,p\hskip-7pt\hbox{$^{^{(\!-\!)}}$} \to W^{\pm} \to l^{\pm} \nu$ and 
$p\,p\hskip-7pt\hbox{$^{^{(\!-\!)}}$} \to Z,\gamma \to l^+ l^-$ ($l=e,\mu$),
and present some numerical results.}{}{}
%\textlineskip			%) USE THIS MEASUREMENT WHEN THERE IS
%\vspace*{12pt}			%) NO SECTION HEADING

\vspace*{1pt}\textlineskip	%) USE THIS MEASUREMENT WHEN THERE IS
\section{Introduction}	%) A SECTION HEADING
\vspace*{-0.5pt}
\noindent
The Standard Model of electroweak interactions (SM)
so far withstood all experimental challenges and
is tested as a quantum field theory  
at the 0.1\% level\cite{Abbaneo:2000nr}. However, the mechanism of
mass generation in the SM predicts 
the existence of a Higgs boson which, so far, has eluded direct
observation. Indirect information on the mass of the Higgs boson, $M_H$,
can be extracted from the $M_H$ dependence of radiative corrections to
the W boson mass. 
Future more precise measurements of the W boson and top quark masses
will considerably improve the present indirect bound on $M_H$:~with 
a precision of 30~MeV for the W boson mass, $M_W$, 
and 2~GeV for the top quark mass which are target values for 
Run~II\cite{prec}, $M_H$
can be predicted with an uncertainty of about $30 \%$.
In order to measure $M_W$ with such precision in a hadron collider
environment it is necessary to fully control higher order QCD and 
electroweak corrections to the W and Z production processes.

\section{{\boldmath Electroweak ${\cal O}(\alpha)$ Corrections to
$p\,p\hskip-7pt\hbox{$^{^{(\!-\!)}}$} \to W^{\pm} \to l^{\pm} \nu$}}
\begin{figure}[htbp]
\vspace*{13pt}
\hspace*{0.5cm}\includegraphics[height=5cm,width=11cm]{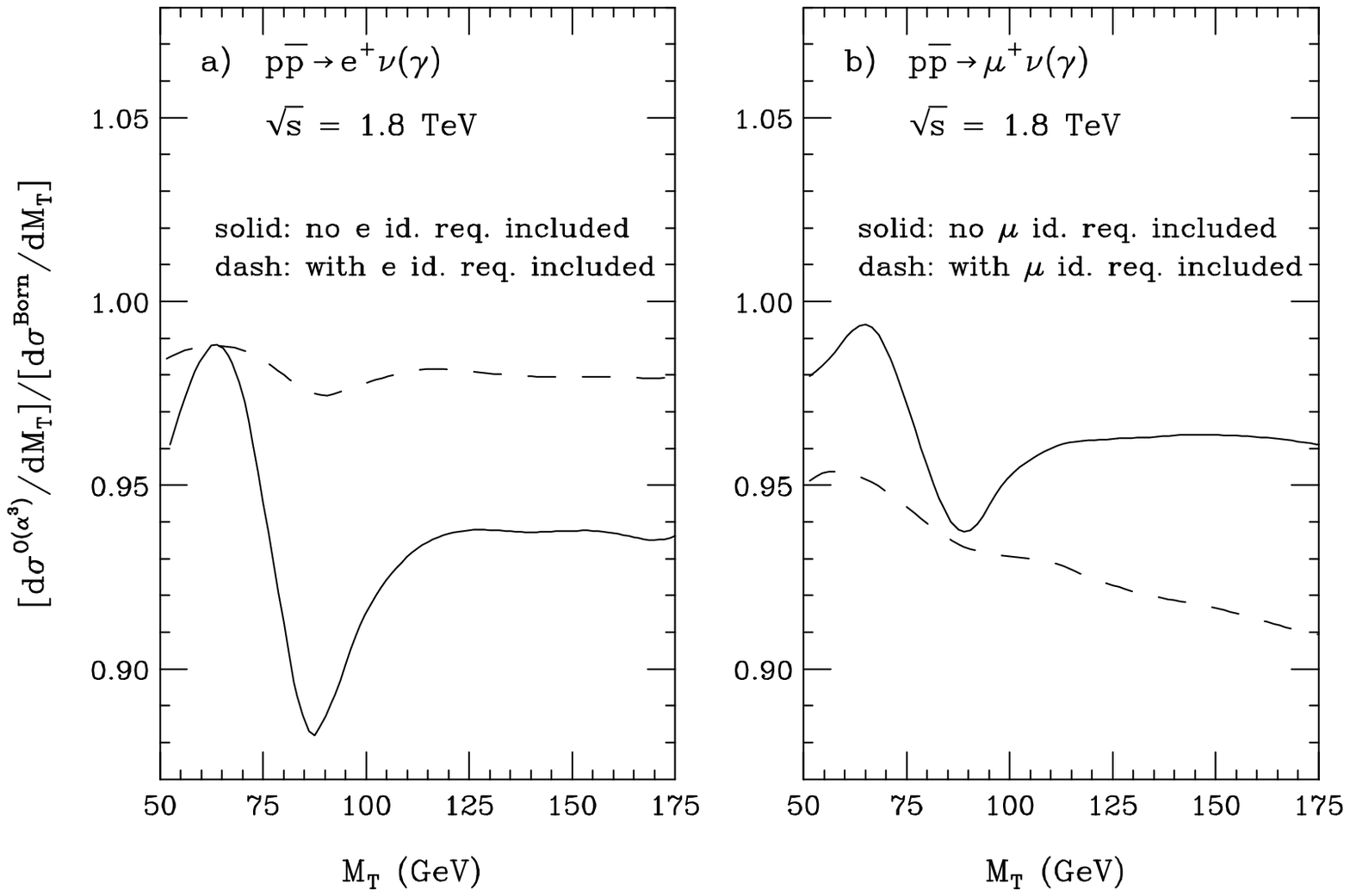}
\vspace*{13pt}
\fcaption{The relative corrections to the $M_T(l \nu)$ distributions
at the Tevatron when only taking into account final state photon radiation
(from Ref.~\ref{Baur1}).}\label{fig:wackone}
\end{figure}

\noindent
The full electroweak ${\cal O}(\alpha)$ 
corrections to resonant W boson production in a general four-fermion
process have been calculated in Ref.~\ref{Wack} with special emphasis
on obtaining a gauge invariant decomposition into a photonic and non-photonic
part. It was shown that the parton cross section for
resonant W boson production via the Drell-Yan mechanism 
$q_i \overline{q}_{i'}\rightarrow f \bar{f}'(\gamma)$ can be written in the 
following form\cite{Baur:1999kt}:
\begin{eqnarray}\label{eqwack:one}
d \hat{\sigma}^{(0+1)} & = &
d \hat \sigma^{(0)}\; [1+ 2 {\cal R}e (\tilde F_{weak}^{initial}(\hat s=M_W^2)+
\tilde F_{weak}^{final}(\hat s=M_W^2))] 
\nonumber \\
&+ & \sum_{a=initial,final,\atop interf.} [d\hat\sigma^{(0)}\; 
F_{Q\!E\!D}^a(\hat s,\hat t)+
d \hat \sigma_{2\rightarrow 3}^a] \; ,
\end{eqnarray}
where the Born cross section, $d \hat \sigma^{(0)}$, is of
Breit-Wigner form and $\hat s$ and $\hat t$ are the usual Mandelstam
variables in the parton center of mass frame.  The (modified) weak
corrections and the virtual and soft photon emission
from the initial and final state fermions (as well as their interference)
are described by the form factors
$\tilde F_{weak}^a$ and $F_{Q\!E\!D}^a$, respectively.  The IR
finite contribution $d\hat\sigma_{2\rightarrow 3}^a$ describes real
photon radiation away from soft singularities.
$F_{Q\!E\!D}^{initial}$ and $d\hat\sigma_{2\rightarrow 3}^{initial}$ 
still include 
quark-mass singularities which need to be extracted and absorbed
into the parton distribution functions (PDFs). 
This can be done in complete analogy to gluon emission in QCD,
thereby introducing a QED factorization scheme dependence.
Explicit expressions for the W production cross section
in the QED DIS and $\overline{\mathrm{MS}}$ scheme are provided in
Ref.~\ref{Baur1}. 
So far, in the extraction of the PDFs from data as well as in the PDF 
evolution, QED corrections are not taken into account.
The latter result in a modified
scale dependence of the PDFs, which is expected to have
a negligible effect on the observable cross sections\cite{Haywood:1999qg}.
The numerical evaluation of the cross section is done with the Monte 
Carlo program
WGRAD\cite{Baur:1999kt}~\footnote{WGRAD is available from the authors.} , and
results have been obtained for a variety of 
interesting W boson observables 
at the Tevatron\cite{Baur:1999kt} and the LHC\cite{Haywood:1999qg}.
In the past, fits to the distribution of 
the transverse mass of the final-state lepton neutrino system, 
$M_T(l\nu)$, have provided the most accurate measurements of $M_W$. 
Photonic initial and initial-final state interference 
corrections were found to have only a small effect on the $M_T$ distribution, 
and weak corrections uniformly reduce the cross section by about $1 \%$.
However, as shown in Fig.~\ref{fig:wackone}, 
final-state photon radiation significantly
distorts the shape of the $M_T$ distribution, and thus
considerably affects the extracted value of $M_W$.  
In the electron case, when taking into account realistic lepton identification 
requirements, the electroweak radiative corrections are strongly reduced
because electron and photon momenta are combined for small opening
angles between the two particles. This eliminates the 
mass singular terms associated with final state radiation.
The correct treatment of the final state radiation has been compared
with a previous approximate calculation\cite{BK} and was found to 
induce an additional shift of ${\cal O}$(10 MeV) in the extracted W
boson~mass\cite{Baur:1999kt}.

\section{{\boldmath Electroweak ${\cal O}(\alpha)$ Corrections to
$p\,p\hskip-7pt\hbox{$^{^{(\!-\!)}}$} \to Z,\gamma \to l^+ l^-$}}
\noindent
Neutral-current Drell-Yan production is interesting for a number of
reasons and it is 
therefore important to determine the electroweak corrections for this
process. Future precise measurements of the W boson mass at
hadron colliders depend on a precise knowledge of the Z boson 
production process. When compared to the values measured at
LEP, the measured Z boson mass
and width help to determine the energy scale
and resolution of the electromagnetic calorimeter. Ratios
of W and Z boson observables may yield a more precise measurement of
$M_W$ than the traditional technique of fitting the $M_T$
distribution\cite{prec}.
The forward-backward asymmetry in the vicinity of the Z
resonance can be used to measure the effective weak mixing 
angle\cite{Haywood:1999qg}.
Finally, at large di-lepton invariant masses, $M_{ll}$,
deviations of the SM prediction could indicate the presence of new physics, 
such as new heavy gauge bosons $Z'$ or extra spatial dimensions.
The QED ${\cal O}(\alpha)$ corrections to 
$p\,p\hskip-7pt\hbox{$^{^{(\!-\!)}}$} \to Z,\gamma \to l^+ l^-$
have been calculated and implemented in the parton level Monte Carlo 
program ZGRAD\cite{Baur:1998wa}~\footnote{ZGRAD is available from the
authors.} . For precision physics away from the Z resonance, however, 
the weak corrections must also be included\cite{Baurinprep}. 
They become important at large values of $M(ll)$ due to the presence
of large Sudakov-like electroweak logarithms of the
form $\ln(M_{ll}/M_Z)$, which 
need to be resummed at very high $M(ll)$. 

Our results show that, for the current level of precision,
the existing calculations for W and Z boson production
are sufficient. However, for future
precision measurements the full electroweak ${\cal O}(\alpha)$ corrections 
and multiple photon radiation effects should be included.
The inclusion of the non-resonant contributions to W production in WGRAD 
is in progress\cite{wbinprep} (see also Ref.~\ref{dprep}).
A calculation of two-photon radiation 
in W and Z boson production at hadron colliders 
has been carried out in Ref.~\ref{Baur2}.

\end{document}